# The $H_0$ tension: did a QCD meV axion emerge?

Massimo Cerdonio[*]
INFN Section and University of Padova, via Marzolo 8, I-35131, Padova, Italy

[*]cerdonio@pd.infn.it

Abstract

I discuss the possibility that the difference in the measured Hubble constant $H_0$ between the current, Late, and the z ~1100, Early, epochs is due to the emergence in between of a new particle. I connect that difference with a change in the effective cosmological constant observed by ΛCDM, which is induced by the energy density of the vacuum of the field of the new particle. Then I try to provide the main characteristics of the particle, boson vs fermion and mass, using the measured change in $H_0$ and a Lorentz invariant regularization of the energy density of vacuum, which relates it to such characteristics. The result indicates that a boson of mass in the range of few meV emerged. A QCD post-inflation cosmological axion in this mass range is allowed by recent analysis and its detection may be attempted according to recent experimental proposals.

1. Introduction

Measurements with various methods of the Hubble constant $H_0$, seen "Late" at low redshifts conflict with the "Early" value given by ΛCDM to levels from 4.0σ up to 5.8σ. The efforts toward a resolution span from observations, with data analysis and interpretation, to theory, with possibly new physics[1].

Here I consider the possibility that the difference in the measured Hubble constant between the z ~ 0, Late, and the z ~1100, Early, epochs is due to the emergence in between of a new particle. I connect that difference with a change in the energy density of vacuum, due to the field of the new particle, which in turn induces the change in the effective cosmological constant $\Lambda_{eff}$ of ΛCDM, implied by the change in $H_0$. Then I find possible to characterize such a putative new particle by looking how mass and statistics can be related to the value of the induced vacuum energy density.

In fact Matt Visser [2] recently discussed at length, quoting all the old and recent literature concerned, how Lorentz invariance gives a definite, finite and cutoff-independent estimate of the zero-point energy contributions of the vacuum of the fields of Standard Model, SM, particles. The estimate is at variance with the one more commonly found still in recent literature, which asks for a high energy cutoff at the Planck scale. This one is seen misleading, in particular not only because it violates Lorentz invariance, but also because it gives the wrong equation of state for the vacuum energy density [3,4]. By contrast the regularization discussed in [2-4], not only is Lorentz invariant, but gives also the correct equation of state for the vacuum [3,4].

2. The proposal

Here I propose a solution to the $H_0$ tension conundrum, by forwarding that between Early and Late epochs a new particle emerged. Quantities will be labeled with "E" and "L", when they refer respectively to the Early and Late epochs and SI units are used.

The regularization of ref [2] provides an estimate for the energy density of the vacuum of the particles, as a sum over (positive) bosonic and (negative) fermionic mass contributions, with logarithmic terms containing a mass scale μ. In the form used in [3,4] for a simplified estimate, it reads as in eq (1) where c is the velocity of light, ℏ the

Planck constant, $m_i$ the mass of the i-particle and $n_i$ the corresponding degrees of freedom

(1) $\rho_v = \pm(c/\hbar)^3 \sum_i n_i (m_i^4/64\pi^2) \ln(m_i/\mu)$

The mass scale $\mu$ needed for regularization must be fixed to get the finite contribution of the particles [3,4]. It is the only parameter not coming in an obvious way from well-established physics. The only indication is that the mass scale $\mu$ must be below the Planck mass scale, because the regularization at that mass scale must be discarded as unphysical, for the reasons noted above. These Authors calculate for the heaviest SM particles an overall (negative) energy density of vacuum $\rho_v \sim -2 \cdot 10^9$ GeV$^4$, when for $\mu$ it is assumed $\mu \sim 3 \cdot 10^{-25}$ GeV. The result is commented to be of little use to compensate an unknown bare $\Lambda_B$ in the attempt to obtain the current effective cosmological constant $\Lambda_{eff}$ of $\Lambda$CDM, as it would be an unacceptable fine tuning [5].

I assume that from eq (1) I can get correctly the *changes* $\Delta\rho_v$ of the vacuum energy density in the case of emergence of a particle with its associated vacuum energy density.

On the other hand, the difference between Late and Early measurements of the Hubble constant implies a difference $\Lambda^L_{eff} - \Lambda^E_{eff}$ in the rate of expansion of the Universe, Late vs Early epochs, as represented in the $\Lambda$CDM model by the observed "effective" cosmological constant $\Lambda_{eff} \sim +1.1 \cdot 10^{-52}$ m$^{-2}$. To connect $\Lambda^L_{eff} - \Lambda^E_{eff}$ to $\Delta\rho_v$, I assume that $\rho_v$ induces in the $\Lambda_{eff}$ a vacuum contribution $\Delta\Lambda_v = (8\pi G/c^2) \Delta\rho_v$. I emphasize that I have no need to enter the issue if the $\Lambda_{eff}$ would come from a combination of a bare cosmological constant $\Lambda_B$ with a $\Lambda_v$, as in the so called semiclassical gravity, or if it originates from some other – yet unknown - mechanism. I only need to keep distinct $\Lambda_B$, should it be there or not, from the $\Lambda_v$ due to the vacuum energy density, and assume of course that $\Lambda_B$, whatever it is, does not suffer any change between the Early and Late epochs. This assumption is reasonable as, according to classical General Relativity, GR, the two constants allowed by the theory - the gravitational G and the cosmological $\Lambda_B$ - are eternal and unchanging since GR emerged. Just as well I never need the total value of $\rho_v$, which can be evaluated from eq (1) for SM particles, but I will use eq (1) to evaluate only *changes* in $\rho_v$ should a new particle be added.

So I take that, as $\Lambda_v$ contributes to the observed $\Lambda_{eff}$, then the changes in $\Lambda_{eff}$ give directly the changes in $\Lambda_v$. The two distinct $H_{0,L}$ and $H_{0,E}$ entail that we have two distinct $\Lambda_{eff}^E$ and $\Lambda_{eff}^L$, and thus two distinct $\Lambda_v^E$ and $\Lambda_v^L$, that is $\Lambda_{eff}^L - \Lambda_{eff}^E = \Lambda_v^L - \Lambda_v^E$.

As $\Lambda_v$ is related to $\rho_v$ as above, the difference $\Lambda_v^L - \Lambda_v^E$ is directly related to the *change* $\Delta\rho_v$ in vacuum energy density, which, according to this proposal was induced by the emergence of the new particle. So from eq (1) I am able to get the main characteristics, boson or fermion and mass $m_X$, of the particle which emerged in between the two cosmic epochs.

According to the spatially flat ΛCDM model the squared Hubble constant as a function of the redshift z, $H^2(z)$, is fittted to get $H[z=0] = H_0$. In this context the radiation contribution is neglected both for the Late and for Early epochs. Attention must be paid to the fact that the Early and Late $H_0$s are fits "anchored" to different ranges of z, which are z ~ 0 (more precisely z <0.01 for Late) and z ~ 1100 for Early, and in fact two distinct $H_{0,L}$ and $H_{0,E}$ are "seen" from the different perspectives .

At any cosmic time t, within the above assumptions, for a flat radiationless Universe [6], it is

(2)   $\Omega_m(t) + \Omega_\Lambda(t) = 1$

where $\Omega_m(t) = [8\pi G/3H^2(t)]\, \rho_m(t)$ and $\Omega_\Lambda(t) = [c^2/3H^2(t)]\Lambda_{eff}$. Writing the evaluation of eq (2) for t=0 – that is z=0 - for the fits coming respectively from the Late and Early epochs, and equating, I find

(3)   $\Delta\rho_{m0} + \Delta\rho_v = (3/8\pi G)\,(H^2_{0,L} - H^2_{0,E})$

where I have inserted $\Delta\rho_{m0} = \rho_{m0}^L - \rho_{m0}^E$ and $\Delta\rho_v = \rho_v^L - \rho_v^E = (c^2/8\pi G)(\Lambda_v^L - \Lambda_v^E)$ where $\rho_{m0}^L$ and $\rho_{m0}^E$ are the mass densities (of all matter) fitted for t=0, as "seen" from the Late and Early epochs respectively. Then eq (3) relates directly the changes in $\rho_{m0}$ and $\rho_v$ to the measured quantities $H_{0,L}$ and $H_{0,E.}$

Despite the $\Delta\rho_v$ is promoted by the $\Delta\rho_{m0}$, and $\Delta\rho_v$ is related to the mass $m_x$ of the putative emerging new particle, no obvious other relation can be written, because the number density of such emerging particle cannot be evaluated in the present context. It is reasonable however to neglect $\Delta\rho_{m0}$ in consideration of the fact that, after the Early emergence of the new particle, it was always small in the expansion from z =1100 to z=0 [7].

I insert in eq (3) the values given by Reiss et al [1] for $H_{0,L}$ and $H_{0,E}$, namely $H_{0,L} = 74$ km s$^{-1}$ Mpc$^{-1}$ and $H_{0,E} = 67.4$ km s$^{-1}$ Mpc$^{-1}$, and I get $\Delta\rho_v = + 1.7\ 10^{-27}$ kg. So, first, as $\Delta\rho_v$ is positive the particle must be a boson. Then eq (1) can be used for the single unknown $m_x$. In fact the other unknown $n_x$, which is already of O(1), would in addition contribute at ¼ power, so I can take $n_x=1$. Using Mathematica I get $m_x \sim 5.3\ 10^{-39}$ kg, that is $m_x \sim 3$ meV. This is obtained for the value of $\mu$ proposed in [4], namely $\mu \sim 3\ 10^{-25}$ GeV. Varying $\mu$ by more than 5 orders around the above value, I get that $m_x$ does not vary more than 10%.

3. Discussion and conclusion.

As said above the regularization mass scale $\mu$ is the only parameter not understood in an obvious way, apart the proposal in [3] taken up again in [4]. However, as seen just above, the $m_x$ comes out to be quite insensitive to the actual value of $\mu$. Still the issue of the method to evaluate $\mu$ remains matter of debate.

On the other hand it is relevant to notice a most recent study [8]. According to that, a rapid evolution of dark energy at z ~1 is strongly indicated by an analysis on available data, and solves the $H_0$ tension (and simultaneously the so called $\sigma_8$ tension not considered here).

A boson in the meV range may well be an axion. Quoting from the Conclusions of ref [9] "In many theoretically appealing ultraviolet completions of the Standard Model axions and axion-like particles occur automatically. Moreover, they are natural cold dark matter candidates. Perhaps the first hints of their existence have already been seen in the anomalous excessive cooling of stars and the anomalous transparency of the Universe for VHE gamma rays."

So on one hand I am allowed to obtain $m_x$ within a context, which concerns the evaluation of the vacuum energy density induced by SM particles only, and on the other hand I find that the value I indicate for $m_x$ is within limits for astrophysical/cosmological axion masses, as given in [9,10]. Papers therein quoted provide predictions for post-inflation QCD axion in a mass ranges up to some 4 meV. In particular one study [11] connects the $H_0$ tension with axions in this mass range. A post-inflation scenario, supported by calculations, is given in [12], and indicates an axion mass range between 0.05 and 1.5 meV.

A panorama of the experimental searches in progress, which would reach also the mass range of interest here, can be found in ref [10] - notice in particular in Figs. 2 and 3 the mass range which will be covered by IAXO. A few more proposals for experiments to detect axions up to the meV range appeared recently in [13,14]. The remarkably new method proposed in [13] would initially work in a mass range, with 0.4 meV as upper limit, but it is foreseen that the method could be extended to larger masses, when appropriate materials will be available.

Acknowledgements
I am much grateful to Alessandro Bettini, Giovanni Carugno and Antonello Ortolan for a critical reading of the manuscript, for useful suggestions towards improving the presentation and for alerting me about papers relevant for the discussion. I gratefully thank Antonio Masiero for relevant comments.

References

[1] L. Verde ,T. Treu and A.G. Riess "Tensions between the Early and the Late Universe" summary of the July 2019 Workshop at Kavli Institute for Theoretical Physics, Santa Barbara, US, in Nature Astronomy **3** 891(2019) arXiv:1907.10625v1; for an (almost) up to date (partial) list of extensions of the ΛCDM cosmology to relieve the tension see refs [17-61] in S.Pan "Reconciling $H_0$ tension in a six parameter space?" arXiv:1907.12551v1; L. Knoxy and M. Millea "The Hubble Hunter's Guide" arXiv:1908.03663v2 discuss a wealth of departures from ΛCDM and find the majority unlikely to be successful; as for theory efforts, a useful summary is given by quoting from A.Riess et al "Milky Way Cepheid Standards for Measuring Cosmic Distances and Application to *Gaia* DR2: Implications for the Hubble Constant" Ap.J. **876** 85(2019):   "…this '$H_0$ Tension' between the early and late Universe…may be interpreted as evidence for a new cosmological feature such as exotic dark energy, a new relativistic particle, dark matter-radiation or neutrino-neutrino interactions, dark matter decay, or a small curvature, each producing a different-sized shift";

[2] Matt Visser   "Lorentz Invariance and the Zero-Point Stress-Energy Tensor" Particles **1** 138 (2018)

[3] J. Martin "Everything you always wanted to know about the cosmological constant problem (but were afraid to ask)"   in *Understanding the Dark Universe*   Comptes Rendus de l'Académie des sciences (Physique) Tome 13 N° 6–7 pag 566 (2012)

[4] J.F.Koksma and T. Prokopec "The Cosmological Constant and Lorentz Invariance of the Vacuum State" arXiv:1105.6296 (2011)

[5] M.Cerdonio "About the *Measure* of the Bare Cosmological Constant" Foundations of Physics **49** 830 (2019), arXiv:1807.08468 using crucially the same regularization, contended that we have rather an indication from well established physics for an *indirect measure* of a primordial bare Lambda, the fine tuning being then offered by Nature; however, apart for the said starting assumption, there is no other relation between this work and the above (see below)

[6] eq (2) is a form of the first Friedman equation for the evolution of a flat Universe dominated by dust matter and an effective cosmological constant, see for instance M.P.Hobson et al. "General Relativity An introduction for Physicists" (Cambridge University Press 2007)

[7] this is comforted by a comment found in A.Riess et al Ap.J. **826** 56 (2016), according to which an uncertainty in $\Omega_m$ of 0.02 induces an uncertainty in $H_0$ of 0.1%

[8] R.E. Keeley, S.Joudaki, M. Kaplinghat and D. Kirkby "Implications of a transition in the dark energy equation of state for the $H_0$ and $\sigma_8$ tensions" arXiv:1905.10198v1

[9] M. Tanabashi et al.(Particle Data Group), Phys. Rev. **D98** 030001 (2018); see: 111. Axions and other similar particles (rev.) page 821

[10] an up to date discussion on astrophysical and cosmological axions, where one finds that axions in a mass range up to few meV (see Fig.3) and even 1 eV(see Fig.2) are comfortably allowed by theoretical and observational constrains, can be found in P.DiVecchia, M.Giannotti, M.Lattanzi and A.Lindner "Round Table on Axions and Axion-like Particles" proceedings of theXIII Quark Confinement and the Hadron Spectrum –Confinement conference arXiv:1902.06567v2; E. Armengaud et al. "Physics potential of the International Axion Observatory " arXiv:1904.09155v3

[11] F.D'Eramo, R. Z. Ferreira, A. Notari and J. L. Bernal "Hot axions and the $H_0$ tension" JCAP 11 (2018) 014 arXiv.org:1808.07430

[12] S. Borsanyi et al. "Calculation of the axion mass based on high-temperature lattice quantum chromodynamics " Nature **539** 69 (2016); M.P.Lombardo "News and Views Particle physics: Axions exposed" Nature **539** 40 (2016)

[13] C. Braggio et al. "Axion dark matter detection by laser induced fluorescence in rare-earth doped materials" Scientific Reports **7** (published online Nov. 9[th] 2017)

[14] M. Lawson, A. J. Millar, M. Pancaldi, E.Vitagliano and F. Wilczek "Tunable Axion Plasma Haloscopes" Phys. Rev. Lett. **123** 141802 (2019